\documentstyle{mn}
\input{epsf.tex}
\newif\ifAMStwofonts

\ifoldfss
  \ifCUPmtlplainloaded \else
    \NewTextAlphabet{textbfit} {cmbxti10} {}
    \NewTextAlphabet{textbfss} {cmssbx10} {}
    \NewMathAlphabet{mathbfit} {cmbxti10} {} 
    \NewMathAlphabet{mathbfss} {cmssbx10} {} 
  \fi
  \ifAMStwofonts
    \ifCUPmtlplainloaded \else
      \NewSymbolFont{upmath} {eurm10}
      \NewSymbolFont{AMSa} {msam10}
      \NewMathSymbol{\upi}     {0}{upmath}{19}
      \NewMathSymbol{\umu}     {0}{upmath}{16}
      \NewMathSymbol{\upartial}{0}{upmath}{40}
      \NewMathSymbol{\leqslant}{3}{AMSa}{36}
      \NewMathSymbol{\geqslant}{3}{AMSa}{3E}

       \let\le=\leqslant
       
    \fi
  \fi
\fi 

\ifnfssone
  \newmathalphabet{\mathit}
  \addtoversion{normal}{\mathit}{cmr}{m}{it}
  \addtoversion{bold}{\mathit}{cmr}{bx}{it}
  \newmathalphabet{\mathbfit} 
  \addtoversion{normal}{\mathbfit}{cmr}{bx}{it}
  \addtoversion{bold}{\mathbfit}{cmr}{bx}{it}
  \newmathalphabet{\mathbfss} 
  \addtoversion{normal}{\mathbfss}{cmss}{bx}{n}
  \addtoversion{bold}{\mathbfss}{cmss}{bx}{n}
  \ifAMStwofonts
    \ifCUPmtlplainloaded \else
      \UseAMStwoboldmath
      \makeatletter
      \new@mathgroup\upmath@group
      \define@mathgroup\mv@normal\upmath@group{eur}{m}{n}
      \define@mathgroup\mv@bold\upmath@group{eur}{b}{n}
      \edef\UPM{\hexnumber\upmath@group}
      \new@mathgroup\amsa@group
      \define@mathgroup\mv@normal\amsa@group{msa}{m}{n}
      \define@mathgroup\mv@bold\amsa@group{msa}{m}{n}
      \edef\AMSa{\hexnumber\amsa@group}
      \makeatother
      \mathchardef\upi="0\UPM19
      \mathchardef\umu="0\UPM16
      \mathchardef\upartial="0\UPM40
      \mathchardef\leqslant="3\AMSa36
      \mathchardef\geqslant="3\AMSa3E

       \let\le=\leqslant

    \fi
  \fi
\fi 

\ifnfsstwo
  \DeclareMathAlphabet{\mathbfit}{OT1}{cmr}{bx}{it}
  \SetMathAlphabet\mathbfit{bold}{OT1}{cmr}{bx}{it}
  \DeclareMathAlphabet{\mathbfss}{OT1}{cmss}{bx}{n}
  \SetMathAlphabet\mathbfss{bold}{OT1}{cmss}{bx}{n}
  \ifAMStwofonts
    \ifCUPmtlplainloaded \else
      \DeclareSymbolFont{UPM}{U}{eur}{m}{n}
      \SetSymbolFont{UPM}{bold}{U}{eur}{b}{n}
      \DeclareSymbolFont{AMSa}{U}{msa}{m}{n}
      \DeclareMathSymbol{\upi}{0}{UPM}{"19}
      \DeclareMathSymbol{\umu}{0}{UPM}{"16}
      \DeclareMathSymbol{\upartial}{0}{UPM}{"40}
      \DeclareMathSymbol{\leqslant}{3}{AMSa}{"36}
      \DeclareMathSymbol{\geqslant}{3}{AMSa}{"3E}

       \let\le=\leqslant

    \fi
  \fi
\fi 

\ifCUPmtlplainloaded \else
  \ifAMStwofonts \else 
    \def\upi{\pi}
    \def\umu{\mu}
    \def\upartial{\partial}
  \fi
\fi
\def\pn{\par\noindent}
\title{Chaotic mixing in noisy Hamiltonian systems}

\author[H. E. Kandrup et al]
  {Henry E. Kandrup,$^{1,2,3}$\thanks{E-mail: kandrup@astro.ufl.edu}
    Ilya V. Pogorelov,$^{2}$\thanks{E-mail: ilya@phys.ufl.edu}
    and Ioannis Sideris$^{1}$\thanks{E-mail: sideris@astro.ufl.edu}\\
   $^{1}$ Department of Astronomy, University of Florida, Gainesville, 
          FL 32611, USA\\
   $^{2}$ Department of Physics, University of Florida, Gainesville, 
          FL 32611, USA\\
   $^{3}$ Institute for Fundamental Theory, University of Florida, Gainesville,
          FL 32611, USA}

\date{Accepted 1999 \hskip 1in .
      Received 1999 \hskip 1in .}

\pagerange{\pageref{firstpage}--\pageref{lastpage}}
\pubyear{1999}

\begin{document}

\maketitle

\label{firstpage}

\begin{abstract}
This paper summarises an investigation of the effects of low amplitude
noise and periodic driving on phase space transport in three-dimensional
Hamiltonian systems, a problem directly applicable to systems like galaxies,
where such perturbations reflect internal irregularities and/or 
a surrounding environment. A new diagnostic tool is exploited to 
quantify the extent to which, over long times, different segments of the same 
chaotic orbit evolved in the absence of such perturbations can exhibit very 
different amount of chaos. First passage times experiments are used to study 
how small perturbations of an individual orbit can dramatically accelerate 
phase space transport, allowing `sticky' chaotic orbits trapped near regular 
islands to become unstuck on surprisingly short time scales. The effects of 
small perturbations are also studied in the context of orbit ensembles with 
the aim of understanding how such irregularities can increase the efficacy of 
chaotic mixing. For both noise and periodic driving, the effect of the 
perturbation scales roughly logarithmically in amplitude. For white noise, the 
details 
are unimportant: additive and multiplicative noise tend to have similar 
effects and the presence or absence of a friction related to the noise by a 
Fluctuation-Dissipation Theorem is largely irrelevant. Allowing for coloured 
noise can significantly decrease the efficacy of the perturbation, but only 
when the autocorrelation time, which vanishes for white noise, becomes so 
large that there is little power at frequencies comparable to the natural 
frequencies of the unperturbed orbit. This suggests strongly that 
noise-induced extrinsic diffusion, like modulational diffusion associated with
periodic driving, is a resonance phenomenon. Potential implications for 
galaxies are discussed.

\end{abstract}

\begin{keywords}
chaos -- galaxies: formation -- galaxies: kinematics and dynamics
\end{keywords}

\section{MOTIVATION}
The objective of the work described here was to explore phase space transport 
in complex Hamiltonian systems that admit both regular and chaotic orbits and, 
especially, to understand how and why low amplitude perturbations, idealised 
as noise and/or periodic driving, can dramatically accelerate the rate at 
which a single chaotic orbit moves from one part of phase space to another.

This work has immediate implications for a variety of problems related to
galactic astronomy. In a first approximation, an elliptical galaxy can perhaps
be characterised by a smooth, time-independent bulk potential. However, this 
is surely not the whole story. One must also allow for discrete substructures, 
e.g., individual stars, and the presence of a surrounding environment,
both of which can have appreciable effects under appropriate circumstances.

It has been long accepted that discreteness effects, i.e., gravitational 
Rutherford scattering between nearby stars, can be modeled as white noise and 
friction by a Fokker-Planck equation. Successive encounters are idealised as a 
sequence of instantaneous kicks associated with random forces that are 
delta-correlated in time, so that the forces at two different instants are 
statistically independent. This noise is then augmented by a dynamical 
friction which represents the systematic drag associated with a star moving 
through the ambient medium. In the context of Chandrasekhar's (1943a) binary 
encounter approximation, the amplitudes of the friction and noise are 
connected by a Fluctuation-Dissipation Theorem (Chandrasekhar 1943b), a 
generic result to be expected in any statistical treatment of an isolated 
system (cf. van Kampen 1981).

Regular motions associated with satellite galaxies and other companion objects 
can trigger a near-periodic driving that will also effect stars in the parent 
galaxy. In terms of the parent mass $M$, the companion mass $m$, the parent 
size $r$, and the characteristic separation $d$ between the galaxies, a 
typical amplitude and frequency are easily estimated. Denoting, respectively, 
by $v$ and $u$ the typical speed of a star in the parent and the relative 
speed of the parent and companion objects, the perturbing force will have an 
amplitude of order $(m/M)(r/d)^{2}$ times as large as the force resulting from 
the bulk potential of the parent galaxy, and the characteristic frequency 
${\omega}$ will scale as ${\omega}{\;}{\sim}{\;}(r/d)(u/v)t_{D}^{-1}$, with 
$t_{D}$ a characteristic dynamical time for the parent galaxy.

A galaxy situated in a high density environment, e.g., a rich cluster, will 
feel a superposition of forces from other nearby galaxies which, generically, 
is far from periodic (although there could be a near-periodic component if, 
e.g., the galaxy has bound satellites). It seems reasonable to suppose that 
these forces are `random,' but it is obviously {\it not} reasonable to 
pretend that they correspond to instantaneous events. 
Idealising them as white noise is clearly inappropriate. However, following 
standard techniques from statistical physics, what does seem appropriate is to 
model them as {\it coloured} noise (cf. Honerkamp 1994), allowing for 
successive impulses which, albeit random, have finite duration. 

Assuming that the noise is a Gaussian random process with zero mean, it is
characterised completely by its second moment, which is assumed to satisfy
\begin{equation}
{\langle}F_{a}(t_{1})F_{b}(t_{2}){\rangle}=K(t_{1}-t_{2}){\delta}_{ab} \qquad
(a,b=x,y,z), 
\end{equation}
with $K$ the autocorrelation function. Demanding that 
$K({\tau})$ be proportional to a Dirac delta ${\delta}_{D}({\tau})$ 
yields white noise, with a flat power spectrum and vanishing autocorrelation 
time. Allowing $K$ to be nonvanishing for a range in ${\tau}$ yields coloured 
noise, with a band-limited power spectrum and a finite autocorrelation 
time. If, consistent with Chandrasekhar's (1941) nearest neighbor 
approximation, one assumes that irregularities associated with the presence of 
other galaxies reflect primarily the effects of a few nearby neighbors, a 
characteristic amplitude and autocorrelation time are again easily estimated. 

Given that the typical distance between galaxies is often less than ten times
the size of a typical galaxy, an external environment can easily induce
perturbations with amplitude at the one percent level or above. The power 
associated with these perturbations should peak at frequencies somewhat smaller
than $t_{D}^{-1}$.

But why might such weak perturbations matter? For example, conventional wisdom 
holds that discreteness effects should be completely irrelevant on time scales 
short compared with the binary relaxation time $t_{R}$, a time which, for a 
galaxy like the Milky Way, is orders of magnitude longer than the age of the 
Universe $t_H$. The answer here lies in the fact that galaxies are more 
complicated dynamical systems than has been appreciated until recently. High 
resolution photometry has provided clear evidence that many galaxies are 
genuinely triaxial, i.e., neither spherical nor axisymmetric, and that most 
galaxies probably have a high density central cusp. Indeed, these features are 
so well established that they have been proposed as the basis of a new 
classification scheme for ellipticals (Kormendy and Bender 1996). However, 
dynamical considerations involving the manifestations of resonance overlap 
provide
compelling reasons to believe that the combination of cusps and triaxiality
leads generically to a bulk potential corresponding to a complex phase space
that admits significant measures of both regular and chaotic orbits (cf.
Merritt 1996, 1999) (although there {\it are} simple examples of cuspy triaxial
potentials which are completely integrable [cf. Sridhar and Touma 1996,1997]).

A chaotic bulk potential can have profound implications for the behaviour of 
stars moving in a galaxy. The sensitive dependence on initial conditions 
characteristic of individual chaotic orbits (cf. Lichtenberg and Lieberman 
1992) implies that initially localised ensembles of chaotic orbits tend to 
diverge exponentially, which leads to a phase mixing far more efficient than 
what obtains for ensembles of regular orbits (Kandrup \& Mahon 1994a, Mahon et 
al 1995, Merritt \& Valluri 1996, Kandrup 1998). This chaotic mixing has 
potentially significant implications for the efficacy with which such 
irregularities as metallicity gradients can disperse in a near-equilibrium 
galaxy. To the extent that galaxies out of equilibrium are dominated by 
chaotic orbits, chaotic mixing could also provide a compelling explanation of 
the obvious efficiency of violent relaxation (Lynden-Bell 1967), as observed, 
e.g., in numerical simulations. 

However, the complex character of the phase space associated with a potential 
admitting both regular and chaotic orbits significantly limits the efficiency 
of chaotic mixing in a time-independent potential. Although the chaotic phase 
space on a constant energy hypersurface is usually connected in the sense that,
in principle, a single orbit can and will eventually access all of it, partial 
obstructions like cantori (cf. Percival 1983) in two-dimensional systems and 
Arnold webs (Arnold 1964) in three-dimensional systems can severely restrict
phase space transport. Although an orbit eventually passes from one phase space
region to another, it can only do so by traversing bottlenecks which so impede 
progress that the time scale for the transit can be thousands of dynamical 
times, this corresponding to a time much longer than $t_{H}$. Indeed, for 
three-dimensional systems one anticipates that, in many cases, the time to 
diffuse along an Arnold web is exponentially long (Nekhoroshev 1977). One 
manifestation of this fact is the so-called `stickiness' phenomenon noted by 
Contopoulos (1971), whereby a chaotic orbit can become stuck near a regular 
island and behave in a near-regular fashion for a time $t{\;}{\gg}{\;}t_{H}$.

The crucial point is that even very low amplitude perturbations
can dramatically accelerate phase space transport through these bottlenecks,
decreasing the time required for an orbit to transit from one phase space 
region to another and, consequently, the time required for an initially
localised ensemble to probe all the accessible phase space. This was first 
recognised in the context of the so-called Fermi acceleration map (Fermi 1949),
a simple symplectic map used to model cosmic ray acceleration, where the 
introduction of white noise was shown to greatly accelerate the rate of 
phase space transport (Lieberman \& Lichtenberg 1972).
More recently, Habib, Kandrup, \& Mahon (1996, 1997) revisited the 
role of white noise in phase space transport, which assumes practical 
importance to accelerator dynamicists concerned with high density charged-beam 
experiments, where discreteness effects can result in the degradation of an 
initially focused beam (Habib \& Ryne 1995). Accelerator dynamicists have also 
recognised that periodic driving can act comparably as a source of accelerated
phase space transport. If orbits are pulsed with a frequency comparable to
their natural frequency, the resulting resonant coupling can dramatically 
alter the rate at which the orbits move through phase space (Lichtenberg \&
Lieberman 1992). 

Low amplitude perturbations may be especially important in near-Hamiltonian
systems like galaxies where the bulk potential is generated self-consistently
by the stars themselves, rather than being imposed externally. Dating back at
least to Binney (1978), galactic dynamicists have assumed that nontrivial
structures like triaxiality in an elliptical or the bar of a spiral cannot
be supported completely by chaotic orbits. Rather, such interesting structures 
would seem to require the presence of various regular orbit families which 
coexist with the chaotic orbits in the complex potential and which, owing to
their shape, can serve as a skeleton. However, several groups interested in
modeling galaxies have found that, because of resonance overlap, there may 
exist almost no regular orbits in crucial phase space regions (e.g., near 
corototation), which might suggest that self-consistent models do not exist.
One proposed solution (cf. Athanassoula et al 1983, Wozniak 1994) has been 
to build the skeleton with `sticky' orbits which, albeit chaotic, can behave 
as if they are nearly regular for times long compared with the age of the 
Universe. Indeed, in the context of numerical modeling based on Schwarzschild's
(1979) method, both Merritt \& Fridman (1996) and Siopis (1998) have concluded 
that near-equilibria appropriate for triaxial Dehnen potentials cannot be 
constructed without including a substantial number of nearly regular chaotic 
segments. 

This seems completely reasonable, but one needs to be sure that the resulting
models are sufficiently stable towards small perturbations reflecting short 
scale irregularities in the potential or the effects of nearby galaxies. The 
problem then is that `sticky' orbits tend to become unstuck much more 
quickly in the presence of low amplitude irregularities than they would in the 
absence of such irregularities. In particular, numerical experiments (Habib, 
Kandrup, \& Mahon 1997) have suggested that extremely weak white noise 
perturbations corresponding to a relaxation time as long as 
$t_{R}{\;}{\sim}{\;}10^{6}-10^{9}t_{D}$ can have major effects within 
${\sim}{\;}100t_{D}$, a time shorter than the Hubble time, even if an 
unperturbed flow remains unchanged for ${\sim}{\;}500t_{D}$ or more. 
Consistent with Merritt (1996), one might therefore anticipate that, in 
response to low amplitude irregularities, galaxies will tend systematically to 
evolve from strongly triaxial configurations towards more nearly axisymmetric
configurations.  

Given the recognition that low amplitude perturbations may play an
important role in the structure and evolution of galaxies, there are 
three obvious questions  to address:
\par\noindent
1. {\it What is the physics that triggers accelerated phase space transport?} 
If nothing else, understanding why this phenomenon arises should help one 
develop an intuition as to when it might prove important.
\par\noindent
2. {\it How does the effect scale with amplitude?} Answering this should
enable one to estimate dimensionally when it is that these effects must be
considered. 
\par\noindent
3. {\it To what extent do the details matter?} Will one get wildly 
different behavior for additive (i.e., state-independent) and multiplicative 
(i.e., state-dependent) noises, or are such state-dependent effects largely
irrelevant? Under what circumstances does allowing for colour, i.e., endowing 
the random forces with a finite autocorrelation time, alter the efficacy of 
the noise? To the extent that the details are unimportant one will have the 
luxury of being able to ignore many complications which are nearly 
inaccessible observationally.

The computations described below suggest strongly that, like periodic driving, 
noise-induced phase space transport is a resonance phenomenon which involves a 
coupling between the frequencies at which the noise has substantial power and 
the natural frequencies of the unperturbed orbit. It follows that allowing for 
a finite autocorrelation time $t_{c}$ has only a minimal effect provided that 
$t_{c}$ remains short compared with $t_{D}$, the natural time scale for the 
unperturbed orbits. However, as $t_{c}$ increases the efficacy of the force 
decreases and, for $t_{c}{\;}{\gg}{\;}t_{D}$, the effects of the noise become
relatively minimal. Overall, the effects scale logarithmically in amplitude. 
The dependence on $t_{c}$ is more complex, but is again better represented as
logarithmic than any simple power law. Other aspects of the perturbation seem
largely immaterial. Additive and multiplicative noises tend to have virtually
identical effects, and the presence or absence of dynamical friction does not
seem to matter appreciably. In this sense, this problem of diffusion through
bottlenecks, so-called entropy barriers (cf. Machta \& Zwanzig 1983), is very
different from energy barrier problems, where the form of the noise tends to
matter a very great deal (cf. Lindenberg \& Seshadri 1981, Alexander \& Habib
1994).

In the past, a number of individuals, including the first author, have 
suggested that noise is important as a source of accelerated phase space
diffusion primarily because it allows one to violate the Hamiltonian 
constraints associated with Liouville's Theorem, which makes hunting through 
phase space easier. This does not seem to be correct. The physics seems to be 
essentially the same as for periodic driving, so it is clearly not the 
non-Hamiltonian character of the noise {\it per se} that is responsible for
what is seen. 

Section II discusses some basic issues related to phase space transport in 
complex time-independent potentials in the absence of all perturbations. 
Section III studies the effects of low amplitude perturbations on individual 
orbits by performing first passage time experiments. What this entails is 
selecting an orbit which is originally trapped near a regular island and 
determining how the characteristic escape time depends on the amplitude and 
form of the perturbation. Section IV focuses on how chaotic mixing is altered 
through the introduction of noise and periodic driving. Section V concludes by 
discussing potential implications for real galaxies.

\section[]{PHASE SPACE TRANSPORT IN CHAOTIC HAMILTONIAN SYSTEMS}

The computations described here were performed for
three-degree-of-freedom Hamiltonian system of the form 
\begin{equation}
H={1\over 2}{\Bigl(}p_{x}^{2}+p_{y}^{2}+p_{z}^{2}{\Bigr)}+V(x,y,z), 
\end{equation}
with $V$ given as a generalisation of
the two-degree-of-freedom dihedral potential (Armbruster et al 1989),
with two free parameters, $a$ and $b$:
$$\hskip -.6in
V(x,y,z)=-(x^{2}+y^{2}+z^{2})+{1\over 4}(x^{2}+y^{2}+z^{2})^{2}$$
\begin{equation}
-{1\over 4}(x^{2}y^{2}+ay^{2}z^{2}+bz^{2}x^{2}), 
\end{equation}
The two-dimensional limit of (3) appropriate for motion with 
$z{\;}{\equiv}{\;}0$ served as a prototypical example in several earlier 
papers (Mahon et al 1995, Habib, Kandrup, and Mahon 1997) which discuss its 
physical 
properties extensively. The fully three-dimensional version was explored in 
the context of chaotic mixing in Kandrup (1998). Note that, for the energies 
used here, $1<E<20$ or so, a dynamical time $t_{D}$ corresponds to 
$t{\;}{\sim}{\;}2-5$, and that most of the power in individual orbits is for 
${\omega}{\;}{\sim}{\;}1-5$.

It is well understood that, because of cantori, chaotic orbit segments in 
two-degree-of-freedom Hamiltonian systems often decompose naturally into two 
(or more) distinct classes, namely (1) unconfined segments which look wildly 
chaotic and tend to avoid regions near regular islands, and (2) confined, or 
sticky, segments which are trapped near regular islands and are nearly
indistinguishable visually from regular orbits. Whether one should expect 
comparable distinctions in three-degree-of-freedom systems is not completely 
obvious. General topological arguments imply that a generic chaotic phase 
space will involve separate regions connected by an Arnold web (Arnold 1964), 
but there is no guarantee that different regions will be particularly 
near-regular or wildly chaotic. Nevertheless, it is clear from visual 
inspection that a single chaotic orbit often does divide naturally into nearly 
regular and wildly chaotic pieces. The obvious question, therefore, is how to
characterise such distinctions in a robust and quantifiable fashion.

One way is to compute estimates of the largest short time Lyapunov exponent 
(cf. Grassberger et al 1988) for different segments of the same chaotic orbit, 
and determine whether $N[{\chi}]$, the distribution of short time Lyapunov 
exponents, exhibits behavior suggestive of multiple populations. Such 
computations were effected here in the usual way by introducing a small 
initial perturbation, evolving the unperturbed and perturbed initial 
conditions, and periodically renormalising the perturbed orbit to ensure that 
the perturbation remains small (cf. Lichtenberg \& Lieberman 1992). Explicitly,
this involved computing an approximation to
\begin{equation}
{\chi}(t)=\lim_{{\delta}Z(0)\to 0}
{1\over t}\ln {|{\delta}Z(t)|\over |{\delta}Z(0)|} 
\end{equation}
with
$|{\delta}Z|^{2}=|{\delta}{\bf r}|^{2}+|{\delta}{\bf p}|^{2}.$ 
Note that, given 
${\chi}(t_{1})$ and ${\chi}(t_{2})$
evaluated in this way, the average exponential instability for the interval 
$t_{1}<t<t_{2}$ satisfies (cf. Kandrup \& Mahon 1994b)
\begin{equation}
{\chi}(t_{2}-t_{1})=
{t_{2}{\chi}(t_{2})-t_{1}{\chi}(t_{1})\over t_{2}-t_{1}}.
\end{equation}

This tact works because chaotic segments which look nearly regular and are 
situated relatively close to regular islands tend systematically to have 
smaller short time Lyapunov exponents than more wildly chaotic segments. (This 
fact, which seems intuitive physically, can be quantified using the notion of 
`orbital complexity' [Kandrup et al 1997], which characterises the extent to 
which an orbit segment has considerable power in a large number of Fourier 
modes. Nearly regular chaotic segments tend to have low complexity, i.e., 
power concentrated in a small number of modes, but they also tend to have 
smaller short time Lyapunov exponents. If short time Lyapunov exponents 
${\chi}$ and complexities $n$ are computed for different segments of the same 
chaotic orbit, one finds typically that, for reasonable sampling times, the 
rank correlation ${\cal R}({\chi},n){\;}{\sim}{\;}0.75-0.95$.)

For a large number of different potentials and energies, data were generated 
by selecting $16$ different initial conditions in the same connected phase
space region and integrating each for a total time $t=524 288=2^{19}$, with
${\chi}(t)$ and the phase space coordinates recorded at intervals 
${\Delta}t=1$. The resulting orbits were partitioned into segments of length 
$t=2^{k}$, with $k$ a positive integer, and short time exponents determined
for each segment using eq.~(5). These short time exponents were then binned
to extract distributions of short time Lyapunov exponents, $N[{\chi}]$, and
the forms of these distributions analysed as functions of $k$ or $\log t$.

Suppose that, for some potential and energy, there is only one `type' of
chaotic orbit, and that the time scale on which the local instability of the
orbit changes significantly, i.e., the autocorrelation time ${\tau}$ for the 
`local stretching numbers,' is short compared with the times over which the 
orbit is being probed. Let ${\delta}t$ denote some basic interval and let 
${\cal N}({\delta}t)$ denote the distribution of short time Lyapunov exponents
for an interval of this length. Assuming only that the moments of 
${\cal N}[{\chi}({\delta}t)]$ exist, the Central Limits Theorem (cf. 
Chandrasekhar 1943b) then makes two specific predictions about longer 
intervals $m{\delta}t$ for $m{\;}{\gg}{\;}1$:
\par\noindent
(1) The longer time distribution $N[{\chi}(t)]$ for $t=m{\delta}t$ will be
approximately Gaussian. 
\par\noindent
(2) The relative width of this Gaussian will decrease as $m^{-1/2}$, so that 
the dispersion ${\sigma}_{\chi}{\;}{\propto}{\;}t^{-1/2}$. 
\par\noindent
Deviations from a Gaussian and/or a dispersion that decreases more slowly are
{\it prima facia} evidence that, over the time scale of in question, the orbit
segments decompose into more than one distinct population.

So what is actually observed? For short times, $t<16-32$ or so, one generally 
sees a singly peaked distribution $N[{\chi}]$. However, this does not 
necessarily imply that there is only one population. For short $t$ the 
distribution $N[{\chi}]$ is so broad that one could easily miss a good deal of 
structure; and it is hard to exclude the possibility of several populations 
with distinct peaks which, when convolved, still yield a unimodal distribution.
As illustrated in the top panels of FIGS. 1 and 2, for small $t$ the 
distribution is typically very smooth and resembles a Gaussian. However, there 
usually {\it are} statistically significant differences from a normal 
distribution. (Panels (a) in FIG. 1 and 2 were each generated by 
binning $2^{23}{\;}{\approx}{\;}8.4\times 10^{6}$ segments, so that even small 
irregularities are significant!) In particular, $N[{\chi}]$ typically has 
a pronounced skew, as might be expected generically if the total distribution 
is comprised of several distinct populations.

As $t$ increases, one sometimes sees evidence for multiple populations, but 
not always. The evidence for multiple populations arises invariably for those 
potentials and energies where visual inspection indicates the possibility of 
trapping near a regular region for reasonably long periods of time. When such 
trapping is {\it not} observed, for $t>32$ or so $N[{\chi}]$ typically 
corresponds very closely to a true Gaussian distribution, with no statistically
significant skew. The smallest values of ${\chi}$ are usually significantly 
larger than zero. Alternatively, when trapping occurs with reasonable 
frequency $N[{\chi}]$ tends instead to reflect the sum of a Gaussian 
distribution centered at a comparatively high value of ${\chi}$ plus one (or 
more) additional lower-${\chi}$ populations which can be manifested as 
contributing a secondary peak to $N[{\chi}]$ and/or an extended low-${\chi}$ 
tail extending down to very small values of ${\chi}$. Examples of potentials 
and energies which exhibit only one and more than one populations are 
exhibited, respectively, in FIGS. 1 and 2.

But how does ${\sigma}_{\chi}$ scale as a function of $t$? When $t$ is 
sufficiently small and the dispersion is large, one finds typically that 
${\sigma}_{\chi}$ decreases as $t^{-1/2}$. This is consistent with the 
existence of only one population, but does not prove that multiple populations 
do not exist: A dispersion ${\sigma}_{\chi}{\;}{\propto}{\;}t^{-p}$ with
$p{\;}{\approx}{\;}1/2$ is also consistent with two distinct populations which,
 however, are
offset by such a small amount as to be completely indistinguishable. For larger
sampling times, more variety is seen. In some cases, coinciding with energies 
and potentials where trapping is at best infrequent, ${\sigma}_{\chi}$ 
continues to scale as $t^{-1/2}$ for larger $t$. However, when trapping is 
more important one finds invariably that, for a finite range of times, the 
dispersion decreases much more slowly than the $t^{-1/2}$ dependence expected 
for a single population. The dispersion of each separate population 
contributing to the total $N[{\chi}(t)]$ may perhaps decrease roughly as 
$t^{-1/2}$, but the composite ${\sigma}_{\chi}$ for $N[{\chi}(t)]$ decreases 
much more slowly. 

This behaviour is illustrated in the upper two panels of FIG. 3 where, for ln 
$t$ between about $3$ and $9$, the total dispersion decreases much more slowly.
If, however, one probes somewhat longer intervals, the distinction between 
populations becomes erased as a single chaotic orbit eventually transits from 
one `type' of chaos to another. At sufficienly late times, there is only one 
population for which, as predicted, ${\sigma}_{\chi}$ decreases as $t^{-1/2}$. 
As illustrated in the two lower panels of FIG. 3, for potentials and energies 
where no evidence for multiple populations is observed, the dispersion is well 
fit by a $t^{-1/2}$ law throughout.

The type of trapping that can arise and how it is manifested by the value of
a short time Lyapunov exponent can be gauged from FIGS. 4 and 5, which were
generated for a single orbit with energy $E=4$ evolved in the potential (2) 
with $a=b=1$. The initial condition was so chosen that, at the outset, the
orbit is wildly chaotic. However, at a time somewhat larger than $t=15 500$
the orbit became trapped near a regular region where it remains stuck for an
interval ${\Delta}t{\;}{\sim}{\;}1000$ before again becoming untrapped, The 
left hand panels
of FIG. 4 exhibit projections of the orbit into the $(x,y)$, $(y,z)$, and
$(z,x)$ planes for the interval $15 700<t<16 212$, an interval ${\Delta}t=
512$ during which the orbit looks nearly regular. The right hand panels exhibit
the same orbit for $15 700<t<16 724$, an interval twice as long. It is clear
that, during the second half of the longer interval, a significant qualitative
change has occured: The orbit no longer manifests the reflection symmetries 
$x\to -x$, $y\to -y$, and $z\to -z$ that were evident during the near-regular 
phase and, as viewed in the $(z,x)$ projection, the orbit is no longer 
centrophobic. Indeed, if the orbit be integrated for a somewhat longer time it 
becomes so wildly chaotic that the three different projections are almost 
indistinguishable visually. 

FIGURE 5, a time series of short time Lyapunov exponents for this orbit, was
generated by computing ${\chi}({\delta}t)$ for successive intervals 
${\delta}t=1$ and, for each data point, performing a box-car average over $256$
adjacent times. At both early and late times ${\chi}$ exhibits considerable
variability. However, these changes are much more modest than what is observed
during the orbit segment's near-regular phase, when ${\chi}$ drops to much 
smaller values. When the exponent only assumes values larger than 
${\chi}{\;}{\approx}{\;}0.10$, the orbit appears visually to be wildly chaotic,
but a drop to values smaller than ${\chi}{\;}{\approx}{\;}0.07$ signals a
transition to a much more regular appearance. 

FIGURE 6 exhibits the analogue of FIG. 4 for a different initial condition,
now evolved with energy $E=10$ in the potential (2) with $a=1.6$ and $b=0.4$.

Three conclusions seems inevitable: (1) As in two-dimensional potentials, it
is possible for chaotic orbits to become trapped near regular regions for very
long times, although they will eventually escape. (2) The presence of such
trapping correlates with the possibility of more complicated phase space 
transport, indicative of the fact that a single connected phase space region
divides into seemingly distinct populations. (3) The existence of {\it de 
facto} populations over some, but not all, finite intervals can be quantified
numerically in terms of the statistics of $N[{\chi}(t)]$, the distribution of
short time Lyapunov exponents.

\section[]{FIRST PASSAGE TIME EXPERIMENTS}

The experiments described here constitute a three-degree-of-freedom 
generalisation of two-degree-of-freedom {\it first passage time} experiments 
in Pogorelov \& Kandrup (1999). These involved identifying
chaotic orbits which, in the absence of any perturbations, remain stuck near
regular islands for very long times, and determining how the introduction of
weak perturbations, idealised as noise or periodic driving, reduces the escape
time. To assess the effects of noise of fixed amplitude and form, large
numbers of noisy integrations of the same initial condition were performed,
each with different pseudo-random seeds. The effects of periodic driving with
frequencies `near' ${\omega}$, involved instead large numbers of integrations
performed with slightly different frequency selected from some small interval 
${\Delta}{\omega}$. The experiments with frequencies $1<{\omega}<100$ 
typically involved uniformly sampling an interval ${\Delta}{\omega}=1$. For 
experiments with $100<{\omega}<1000$, the interval was ${\Delta}{\omega}=10$; 
for those with $0.1<{\omega}<1.0$, ${\Delta}{\omega}=0.1$. Each set of 
computations involved between $1000$ and $5000$ orbits.

In the absence of any perturbations, determining precisely the phase space 
regions corresponding to a `sticky' chaotic orbit is possible, albeit very
tedious (Contopoulos, private communication). However, the notion of a well
defined boundary necessarily evaporates as soon as the orbits are perturbed:
because of the perturbations, energy is no longer conserved, so that the 
effective phase space hypersurface on which the orbit
moves changes continually as the system evolves. For this reason, `escape'
was defined in a more practical fashion. Specifically, simple polynomial 
formulae were used to delineate approximately the configuration space region
to which the orbit was originally confined, and the first escape time was 
identified as the first time the orbit leaves this region. That the escape
criterion is reasonable was tested in two important ways: It was verified that,
with or without perturbations, small changes in the boundary change the escape
time only minimally; and that, at least in the
absence of perturbations, `escape' coincides with an abrupt increase in the
short time Lyapunov exponent. 

In analysing the data, two simple diagnostics proved especially convenient:
\pn
1. {\it The time $T(0.01)$ required for one percent of the orbits in the 
computation to escape.} As described below, escape does not begin immediately.
Rather, there is typically an extended initial period, the duration of which
depends on the form and amplitude of the perturbation, during which there are
no escapes. Escapes then turn on abruptly, at a time well characterised overall
by $T(0.01)$. ($T(0.01)$ is less sensitive to statistical fluctuations than
$T_{0}$, the time the first orbit escapes.)
\pn
2. {\it The initial escape rate ${\Lambda}$.} In most, albeit not all, cases 
escapes, once they begin, appear to sample a Poisson process, at least 
initially, so that $N(t)$, the number of orbits not yet having escaped, 
decreases exponentially. 
\pn
A good representation for the data, used in the analysis, was 
\begin{equation}
N(t)=N_{0}\exp {\Bigl\{}-{\Lambda}{\bigl[}t-T(0.01){\bigr]}{\Bigr\}} 
\hskip .12in {\rm for}\;\;t>T(0.01).
\end{equation}

The experiments with periodic driving entailed solving an evolution equation
of the form
\begin{equation}
{d^{2}{\bf r}\over dt^{2}}=-{\nabla}V({\bf r}) + 
A \sin\,{\omega}t\;{\hat{\bf r}}
\end{equation}
for variable $A$ and ${\omega}$  Those with noise involved solving
Langevin equations (cf. Chandrasekhar 1943b, van Kampen 1981)
\begin{equation}
{d^{2}{\bf r}\over dt^{2}}=-{\nabla}V({\bf r}) - {\eta}{\bf v} + {\bf F}.
\end{equation}
Here ${\eta}={\eta}({\bf v})$ is the coefficient of dynamical friction
and the stochastic force {\bf F} is homogeneous Gaussian noise, with moments
$$\hskip -1.6in
{\langle}F_{a}(t){\rangle}=0 \qquad (a,b=x,y,z) $$
and
\begin{equation}
{\langle}F_{a}(t_{1})F_{b}(t_{2}){\rangle}=K({\bf v},t_{1}-t_{2}){\delta}_{ab}.
\end{equation}
For delta-correlated white noise, 
\begin{equation}
K({\bf v},{\tau})=2{\Theta}{\eta}({\bf v}){\delta}_{D}({\tau}),
\end{equation}
with ${\Theta}$ a characteristic `temperature,' i.e., a typical
energy for the internal degrees of freedom responsible for the noise. Coloured
noise replaces ${\delta}_{D}({\tau})$ by a function of finite width. Most of
the integrations involved choosing ${\Theta}{\;}{\sim}{\;}E$.

Internal irregularities should give rise to both friction and noise, so that 
it is not reasonable physically to consider one without the other. However, 
the relative importance of friction versus noise as a source of phase space 
transport was tested by comparing computations including both friction and 
noise with computations with the friction ${\eta}{\bf v}$ turned off. 
The white noise simulations were performed using an algorithm developed by
Greiner et al (1988) [see also Honerkamp (1994)]. The coloured noise 
simulations
were implemented using a new algorithm developed by I.~V.~Pogorelov as part of 
his Ph.~D. dissertation. The basic idea is outlined in Pogorelov \& Kandrup 
(1999). 

\subsection{The effects of periodic driving}

As noted already, escape is a two stage process. At very early times there are
no escapes, and the only significant effect of the driving is to cause 
different integrations of the same initial condition with different frequencies
to diverge exponentially inside the trapped region. Overall, for a fixed
frequency interval ${\Delta}{\omega}$, the {\it rms} dispersion in the 
separation of different orbits scales as 
\begin{equation}
{\delta}r_{rms}{\;}{\propto}{\;}A\,\exp({\chi}t),
\end{equation}
where $A$ is the amplitude of the driving and ${\chi}$ is comparable to a 
typical value for the largest short time Lyapunov exponent. Only after the
orbit ensemble has dispersed to probe significant portions of the trapped
region and ${\delta}r_{rms}$ approaches some critical value do any escapes 
occur. These turn on quite abruptly, the interval during which (say) the first 
percent of the orbits escape being much shorter than the time before the first 
orbit escapes. That escapes begin when ${\delta}r_{rms}$ approaches some 
critical value implies that the time at which escapes begin, as probed, e.g., 
by the one percent escape time $T(0.01)$, scales logarithmically in amplitude. 

The earlier stages of the actual escape process, during which 90 percent or so
of the orbits escape, are typically well represented by a Poisson process,
with $N(t)$, the fraction of the orbits that have not yet escaped, decreasing 
exponentially. This allows for the simple interpretation that, once the orbits
have diverged to probe all the trapped regions, they can and will escape `at
random' as they find an appropriate exit channel. However, one finds typically
that, at later times, escapes proceed more slowly, so that the decrease
in $N(t)$ becomes subexponential. The reason for this is not completely
clear, but it seems reasonable to suppose that some of the orbits that did
not escape early on became trapped even more closely to the regular region,
so that escape is more difficult than initially. (Because energy is no longer
conserved, it is possible for an initial condition which, in the absence of 
irregularities, corresponds to a chaotic orbit to move into a phase space
region where, in the absence of the irregularities, it would be regular, at 
which point escape becomes exceedingly difficult.) For very small
amplitudes $A$, the escape rate ${\Lambda}$ is often relatively insensitive to 
$A$, which suggests that the driving does not facilitate escapes all that much.
However, for larger amplitudes, say $A>10^{-6}-10^{-5}$ one typically finds
that ${\Lambda}$ exhibits a roughly logarithmic dependence on $A$. Increasing
the amplitude increases the escape rate. This indicates that the driving
helps the orbits access escape channels by jiggling them about.

Periodic driving has the strongest effect for frequencies ${\omega}
{\;}{\sim}{\;}t_{D}^{-1}$, which allow an efficient resonant coupling
between the driving frequency and the natural frequencies of the unperturbed
orbit. However, periodic driving can still have a significant effect at much 
larger and much smaller frequencies, presuming by couplings through harmonics. 
Even frequencies as large as ${\omega}=1000$ can significantly reduce the 
escape time. A detailed plot of $T(0.01)$ or ${\Lambda}$ as
a function of ${\omega}$ exhibits considerable structure but, overall, the
dependence on frequency is roughly logarithmic.

FIGURE 7 exhibits representative plots of $\ln\,N(t)$, generated for different
integrations of the same initial condition with $E=10$ evolved in the 
potential (2) with $a=0.4$ and $b=1.6$. Each ensemble of $1000$ orbits was
generated by freezing the amplitude at a fixed value $A$ and uniformly 
sampling the interval $5<{\omega}<6$ with $1000$ different driving frequencies.
The different curves correspond to different amplitudes. It is clear that,
in each case, there is a finite initial interval during which there 
are no escapes, followed by an interval during which $N(t)$ exhibits a 
roughly exponential decrease. The amplitude dependence of the escape process 
is illustrated in FIGS. 8 (a) and (b), which exhibit $T(0.01)$ for the 
same initial condition for two different frequency intervals, $2<{\omega}<3$
and $5<{\omega}<6$. FIGS. 8 (c) and (d) exhibit the frequency dependence
of $T(0.01)$, as extracted from a collection of simulations with fixed 
amplitude $A=10^{-3}$. 

\subsection{The effects of white noise}

The effects of white noise are very similar to those of periodic driving, a
fact that can be understood if one recognises that, in a real sense, white
noise is an incoherent sum of oscillations with all possible frequencies. 
Gaussian white noise, with random phases and a flat power spectrum, is 
equivalent mathematically to a superposition of periodic oscillations with all 
possible frequencies and all possible phases!

Overall, the effects of noise are again manifested as a two-stage process:
a phase during which different orbits in the ensemble -- now different noisy 
realisations of the same initial condition evolved to sample the same random 
process -- disperse within the confining boundary, followed by a period of 
escapes reasonably well approximated as a Poisson process.
FIG.~9 illustrates the typical behaviour of $\ln N(t)$ allowing for a constant
coefficient of dynamical friction and additive noise, generated for the same 
initial condition used to create FIG. 7. The different curves have the same
``temperature,'' ${\Theta}=E=10.0$, but different values of ${\eta}$. As for 
the case of periodic driving, $T(0.01)$ increases logarithmically with 
decreasing amplitude -- now measured by ${\eta}$ -- at least for ${\eta}$
not too small, say ${\eta}>10^{-9}-10^{-8}$. The slope ${\Lambda}$ 
associated with the exponential decay of $N$ also tends to decrease with 
decreasing ${\eta}$, which indicates that, even after escapes have begun, 
noise accelerates the escape process by jiggling orbits and thus helping them 
to find an escape channel. FIGS.~10 (a) and (b) exhibit $T(0.01)$ as a 
function of $\log \eta$ for two different initial conditions. 

As for the case of periodic driving, the effects of the noise largely
disappear when the amplitude becomes too small, ${\eta}<10^{-10}-10^{-9}$ or 
so. This can be understood at least in part as a numerical artifact. 
The white noise simulations were effected using a fixed ${\delta}t$ time 
step, fourth order Runge-Kutta integrator, which leads to ``random'' errors of 
order $({\delta}t)^{4}$. Given a time step ${\delta}t=10^{-3}$, the 
computations might be expected to incorporate ``numerical noise'' of amplitude 
${\sim}{\;}10^{-10}$, and additional irregularities of lower amplitude
should have only minimal effects.

Another significant conclusion is that both the qualitative and quantitative 
effects of noise seem comparatively insensitive to the details. Allowing for
at least some simple forms of multiplicative noise and/or allowing for a 
variable coefficient of dynamical friction or turning off the friction
altogether has only a minimal effect. As an illustrative example, one can 
consider what happens if the quantity ${\eta}$ entering into both the friction 
and the autocorrelation function becomes a nontrivial function of velocity,  
assuming ${\eta}{\;}{\propto}{\;}v^{p}$. The basic conclusion of such an 
investigation, illustrated by FIG.~11, is that these changes have almost no
effect. Here the solid curve represents additive noise and a constant 
coefficient of dynamical friction, ${\eta}_{0}$. The dot-dashed, 
triple-dot-dashed, and dotted curves represent, respectively, friction and 
noise with 
${\eta}={\eta}_{0}v^{p}/{\langle}v^{p}{\rangle}$, with $p=1$, $2$, and $3$, and
${\langle}v^{p}{\rangle}$ a mean value computed for the unperturbed orbit. 
(This normalisation ensures that the ``average'' noise is the same for the 
additive and multipicative simulations.) The dashed curve 
corresponds to additive white noise but vanishing friction (i.e., allowing for
a nonzero ${\eta}_{0}$ in the autocorrelation function but assuming a vanishing
coefficient of dynamical friction). The obvious point is that, early on, these 
curves are virtually identical and that, even at later times, the differences 
are comparatively minimal. 

In principle, one could perhaps artificially ``tune'' the form of the noise 
to enhance or suppress its effects, e.g., by making kicks especially large when
an orbit approaches an escape channel. However, this seems unphysical. To the 
extent that the ``noise'' impacting real stars is largely uncorrelated 
with the physics of the bulk potential, which determines the locations of 
these escape channels, one might anticipate that the details will be 
comparatively unimportant.

\subsection{The effects of coloured noise}

The aim of the work described here was to determine how the aforementioned
results regarding friction and white noise are altered if the noise becomes
coloured. In other words, what happens if the autocorrelation function 
$K({\tau})$ is not delta-correlated in time, so that the random impulses to
which the orbits are subjected are of finite duration?

Two comparatively simple examples were considered. The first corresponds to
the so-called Ornstein-Uhlenbeck process (cf. van Kampen 1981), for which 
$K({\tau})$ decays exponentially:
\begin{equation}
K({\tau})={\alpha}{\eta}{\Theta}\exp(-{\alpha}|{\tau}|).
\end{equation}
The second alternative involved an exponential modulated by a power law:
\begin{equation}
K({\tau})={3{\alpha}{\eta}{\Theta}\over 8}\exp(-{\alpha}|{\tau}|)
{\Bigl(} 1 + {\alpha}|{\tau}| + {{\alpha}^{2}\over 3}{\tau}^{2}{\Bigr)}.
\end{equation}
In each case, the normalisations were so chosen so that
\begin{equation}
\int_{-\infty}^{\infty}\,K({\tau})d{\tau}=2{\Theta}{\eta}.
\end{equation}
In other words, fixed noise amplitude means a fixed value $2{\Theta}{\eta}$ 
for the time integral of the autocorrelation function (which equals the
diffusion constant $D$ entering into a Fokker-Planck description [cf.
Chandrasekhar 1943b, van Kampen 1981]). The autocorrelation 
times for these processes are, respectively, $t_{c}=1/{\alpha}$ and 
$t_{c}=2/{\alpha}$. The white noise calculations described in the preceding 
section can be understood as involving a singular limit ${\alpha}\to\infty$. 
As for the case of multiplicative noise, these two examples only probe the 
tip of an iceberg. However, an analysis of their effects 
{\it does} provide insight into the question of how a finite autocorrelation 
time can impact phase space transport in a complex phase space.

As for white noise, the evolution of an ensemble of orbits in the presence of
coloured noise is a two stage process. After an initial interval without
escapes, during which different members of the ensemble diverge exponentially,
escapes turn on abruptly, with the first percent of the orbits escaping within
a time $T(0.01)$ much shorter than the time before the first escape. This is
then followed by a phase during which $N(t)$ decreases systematically. As for
the case of white noise, this phase can usually be well fit overall by an
exponential, although a plot of $N(t)$ for coloured noise tends to exhibit a 
bit more structure than do comparable plots for white noise or periodic 
driving. For fixed autocorrelation time $t_{c}$, $T(0.01)$ exhibits a roughly 
logarithmic dependence on ${\eta}$, at least for $10^{-8}<{\eta}<10^{-4}$ or 
so; and the slope ${\Lambda}$ associated with the exponential decrease in
$N(t)$ tends to scale logarithmically with ${\eta}$. One also finds that the
presence or absence of friction is largely immaterial.

In all this, coloured noise behaves just like white noise. The real question
is: how do things depend on $t_{c}$? When $t_{c}$ is very small and ${\alpha}$
is very large, the effects are nearly indistinguishable from white noise:
both $T(0.01)$ and ${\Lambda}$ are essentially unchanged. Significant 
deviations only begin to arise when the autocorrelation time $t_{c}$ becomes
comparable to the dynamical time $t_{D}$ for the unperturbed orbit. At this
stage, the noise begins to have an appreciably weaker effect, its overall 
efficacy scaling logarithmically in ${\alpha}$ or $t_{c}$. 

FIGURE 12 exhibits $\ln\,N(t)$ for a fixed ensemble evolved with coloured noise
satisfying eq. (13) for ${\Theta}=10$ and ${\eta}=10^{-5}$, but allowing 
for several different values of ${\alpha}$. FIGURE 13 shows examples of how 
$T(0.01)$ scales with ${\eta}$ for fixed nonzero ${\alpha}$ (a and c) and how 
$T(0.01)$ scales with ${\alpha}$ for fixed ${\eta}$ (b and d). 

The observed behaviour is easy to understand. Increasing $t_{c}$ from zero to 
a finite
value is equivalent to replacing the flat white noise power spectrum by a 
band-limited power spectrum. Deviations from white noise begin to be important
when this spectrum is so limited that there is comparatively little power at
frequencies comparable to the natural frequencies of the unperturbed orbit. In
this sense, it would appear that, like modulational diffusion associated with 
periodic driving, noise-induced extrinsic diffusion should be interpreted as
a resonance phenomenon. The fact that the efficacy of the perturbation scales
logarithmically in ${\alpha}$ is reminiscent of the fact that, for periodic
driving, the efficacy scales logarithmically in driving frequency.

\section[]{CHAOTIC MIXING IN THE PRESENCE OF PERIODIC DRIVING AND NOISE}
The aim of the work summarised in this Section was to determine how chaotic 
mixing is impacted by low amplitude perturbations idealised as periodic 
driving or friction and noise. This entailed selecting localised ensembles of 
initial conditions and tracking their behaviour as they are integrated into 
the future, both with and without perturbations. The resulting orbital data 
were analysed as in Kandrup (1998) by computing both (i) coarse-grained 
representations of reduced distribution functions $f(Z_{a},Z_{b})$ for 
different pairs of phase space coordinates and (ii) time-dependent moments 
${\langle}x^{i}y^{j}z^{k}p_{x}^{l}p_{y}^{m}p_{z}^{n}{\rangle}$ for
$i+j+k+l+m+n{\;}{\le}{\;}4$.

The principal conclusion is that low amplitude perturbations can impact the
evolution in two potentially significant ways:

Even in settings where ``stickiness'' is not important, so that orbits can
spread out with comparatively few obstacles, time-dependent perturbations can
play a role in damping oscillations and ``fuzzing out'' short wavelength 
structures. Because of the exponential sensitivity generic for chaotic systems,
a localised ensemble of initial conditions corresponding to ``unconfined''
chaotic orbits will, in the absence of any perturbations, {\it diverge} 
exponentially. However, this tendency of nearby orbits to diverge implies 
that the dispersing ensemble will eventually {\it converge} towards an 
invariant, or near-invariant, distribution, i.e., a (near-)equilibrium 
(Kandrup 1998; see also Merritt \& Valluri 1996). If, e.g., the orbital data 
be binned at succesive instants to generate a gridded representation of some 
$f(Z_{a},Z_{b},t)$, one finds that, with respect to a discrete $L^{1}$ or 
$L^{2}$ norm, this time-dependent $f(t)$ will typically evolve exponentially 
towards some nearly time-independent $f_{niv}(Z_{a},Z_{b})$, i.e.,
\begin{equation}
{\Bigl(}
{\sum_{a}\sum_{b}|f(Z_{a},Z_{b},t)-f_{niv}(Z_{a},Z_{b})|^{p} }{\Bigr)}^{1/p}
{\;}{\sim}{\;}e^{-{\Lambda}t}, 
\end{equation}
for $p=1$ or $2$. Similarly, moments like ${\langle}x{\rangle}$ or
${\langle}xp_{y}{\rangle}$, which vanish for an invariant distribution, 
evolve towards zero exponentially.

However, this evolution is not always uniform in time. In many cases, the 
systematic evolution is accompanied by coherent oscillations that only damp on 
comparatively long time scales. For example, a plot of the dispersion in some 
phase space variable can exhibit significant oscillations that persist for 
tens of dynamical time $t_{D}$ even if the ``average'' dispersion has settled 
very nearly to its equilibrium value within a time $t<10t_{D}$. Allowing for 
noise and/or periodic driving can significantly decrease the amplitude of 
these oscillations. 

This behaviour is illustrated in FIG. 14 (a) and (c) which track ${\sigma}_{x}$
for the same ensemble of initial conditions evolved in the potential (b) with 
$a=b=1$ both with and without periodic driving. In each cases, the dispersion 
converges towards a value ${\sigma}_{x}{\;}{\approx}{\;}1.2$, but for the 
perturbed orbits the convergence is clearly more efficient. 
FIGS.~15 (a) and (c) track ${\sigma}_{x}$ for the same ensemble, now evolved
in the presence of additive white noise with ${\Theta}=E$ and, respectively, 
${\eta}=10^{-4}$ and ${\eta}=10^{-6}$. It is clear that weak noise 
can damp irregularities even more efficiently than can periodic driving.

The second important point is that, because of the escape phenomenon described 
in the preceding section, weak perturbations can accelerate the diffusion
of sticky orbits, allowing them to probe phase space regions which otherwise 
are only accessed at comparatively late times. For example, as noted in 
Kandrup (1998), ensembles of chaotic orbits with small initial $z$ and $p_{z}$ 
tend, in the absence of perturbations, to remain relatively close to the 
$z$-axis, so that they only approach an invariant distribution on a relatively 
long time scale -- $t{\;}{\sim}{\;}50$ rather than $t{\;}{\sim}{\;}5$. However,
low amplitude perturbations can reduce this time scale significantly by 
allowing these orbits to move away from the $z$-axis, thus enabling them to 
probe all (or at least a signficantly larger portion of) the connected phase 
space region. 

This is illustrated in FIGS. 14 (b) and (d), which track ${\sigma}_{z}$ for
the same ensemble used to generate FICS. 14 (a) and (c). Here once again the
top panel represents an unperturbed orbital integration whereas the lower panel
is generated from a simulation subjected to a finite amplitude periodic 
driving. In interpreting the plots of ${\sigma}_{x}$ and ${\sigma}_{z}$, it 
should be noted that the initial ensemble was chosen so that $x=0.0$, $z=0.2$,
and $p_{z}=0.0$. The initial $y$ and $p_{y}$ were generated by uniformly
sampling, respectively, the intervals $[0.8,1.0]$ and $[2.4,2.6]$ and $p_{x}>0$
was computed as a function of the remaining phase space variables to ensure 
that $E=10$. The unperturbed dispersion ${\sigma}_{z}$ in FIG. 14 (b) is 
typical of what is seen when evolving a collection of relatively sticky orbits.
FIG.~14 (d) more closely resembles what one would have expected to see for an 
unperturbed evolution of orbits that are {\it not} initially trapped in a 
special phase space region. FIGS.~15 (b) and (d) exhibit analogous plots of
${\sigma}_{z}$ generated from the noisy simulations used to generate FIGS.~15
(a) and (c). The trend is again the same, although the effect is not quite as
pronounced

What this could imply for the visual appearance of a galaxy is illustrated in 
FIGS. 16 and 17, which exhibits grey scale plots generated from the same orbit 
integrations. The first panel in FIG. 16 exhibits the distribution $f(z,y)$ 
for an unperturbed integration, generated from data recorded at
intervals ${\delta}t=1$ for $64{\;}{\le}{\;}t{\;}{\le}{\;}128$. The obvious
point is that, even though the potential exhibits cubic symmetry, the orbits
are more localised in the $z$-direction than in the $y$- (or $x-$) direction.
The second panel exhibits $f(z,y)$ much later on, namely for 
$896{\;}{\le}{\;}t{\;}{\le}{\;}1024$, by which time the distribution has become
much more symmetric. The lower panels, again generated for 
$64{\;}{\le}{\;}t{\;}{\le}{\;}128$, exhibit $f(z,y)$ for the same initial 
ensemble now evolved in the presence of periodic driving with two different
frequencies. FIGS. 17 (a) and (b) exhibit analogous plots generated from 
noisy integrations, It is evident in each case that, because of the 
perturbations, the distribution is more nearly symmetric than was the case for 
an integration with no perturbations.

\section[]{DISCUSSION}

To assess the importance of all this for galactic dynamics, one must 
determine the extent to which orbit trapping and slow phase space diffusion, 
seemingly generic phenomena for nonintegrable potentials that admit a 
coexistence of both regular and chaotic orbits, are present in the
cuspy, traxial potentials which appear to characterise many galaxies.

As a concrete example, consider the maximally triaxial ${\gamma}=1$ 
Dehnen potential (Merritt and Fridman 1996), which is generated 
self-consistently from the mass density 
\begin{equation}
{\rho}(m)={(3-{\gamma})M\over 4{\pi}abc}m^{-{\gamma}}(1+m)^{-(4-{\gamma})}
\end{equation}
with
\begin{equation}
m^{2}={x^{2}\over a^{2}}+{y^{2}\over b^{2}}+{z^{2}\over c^{2}}, 
\end{equation}
for ${\gamma}=1$, $c/a=1/2$, and $(a^{2}-b^{2})/(a^{2}-c^{2})=1/2$. 
For each of twenty different energies, some $300-400$ well separated initial 
conditions were identified, all corresponding to chaotic orbits. [These
were the chaotic initial conditions used to generate Siopis's (1998) library 
of orbits for the construction of equilibria using Schwarzschild's (1979) 
method.] Each initial conditions was integrated for a time $2048t_{D}$, with 
${\chi}(t)$ recorded at intervals ${\delta}t=1$. As for the computations 
described in Section 2, the resulting data were then partitioned into segments 
of various length $t$ and the dispersion ${\sigma}_{\chi}$ was studied as a 
function of $t$. 

The principal conclusion from this analysis is that, at least for this
potential, unperturbed phase space transport always proceeds {\it very} slowly.
For relatively high energies, the observed distributions of short time Lyapunov
exponents are similar to what is found for the potential (2) for choices of
the parameters $a$ and $b$ and the energy $E$ for which ``stickiness'' is a
frequent occurence. Moreover, as for that potential, the time scale on which 
details wash out and the dispersion begins to decay as $t^{-1/2}$ is of
order $1000t_{D}$. However, for the lowest energies, where chaotic orbits
pass very close to the central cusp, phase space transport appears to proceed
even more slowly! Distributions of short time Lyapunov exponents tend to
exhibit considerable structure, even for comparatively long times. Indeed, 
even after $2048t_{D}$ there is no suggestion that the distinctions between 
different populations have been erased. This is illustrated in FIG. 18, which 
exhibits ${\sigma}_{\chi}(t)$ for the two highest and lowest energies used by 
Merritt \& Fridman (1996) in their construction of Schwarzschild equilibria.

The obvious inference is that, at least for this triaxial Dehnen potential,
unperturbed chaotic orbits often diffuse through phase space only very slowly,
especially at low energies. The two obvious questions then are: (i) is this 
result generic for cuspy triaxial potentials, or just an accident for this
specific potential, and (ii) to what extent can low amplitude friction and
noise accelerate even this very slow phase space transport? Both these 
questions are currently being addressed (Siopis \& Kandrup 1999).

\section*{Acknowledgments}
It is pleasure to acknowledge useful discussions with Christos Siopis.
Partial financial support was provided by the Institute for Geophysics and 
Planetary Physics at Los Alamos National Laboratory. The simulations involving 
coloured noise were performed using computational facilities provided by Los 
Alamos National Laboratory. Work on this manuscript was completed while HEK
was a visitor at the Aspen Center for Physics, the hospitality of which is
acknowledged gratefully.

\vfill\eject
\pagestyle{empty}
\begin{figure}[t]
\centering
\centerline{
        \epsfxsize=8cm
        \epsffile{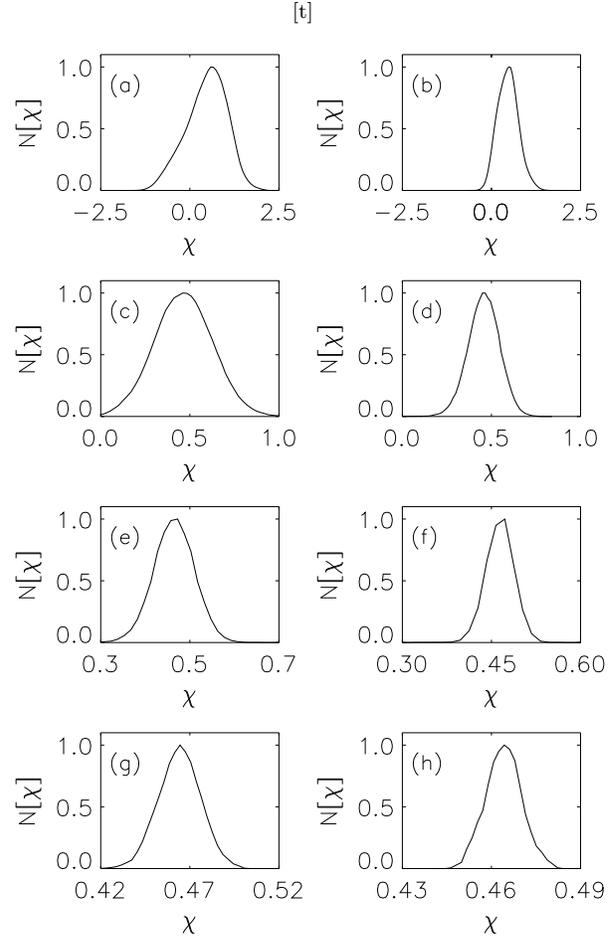}
           }
        \begin{minipage}{10cm}
        \end{minipage}
        \vskip -0.0in\hskip -0.0in
\caption{ $N[{\chi}(t)]$, the distribution of short time Lyapunov exponents
for a collection of chaotic orbits with energy $E=6$ evolved in the potential
(2) with $a=3$ and $b=-1$, computed as a function of the sampling time $t$.
(a) $t=1$. (b) $t=4$. (c) $t=16$. (d) $t=64$. (e) $t=256$. (f) $t=1024$.
(g) $t=4096$ (h) $t=16384$.}
\vspace{0.0cm}
\end{figure}
\vfill\eject

\pagestyle{empty}
\begin{figure}[t]
\centering
\centerline{
        \epsfxsize=8cm
        \epsffile{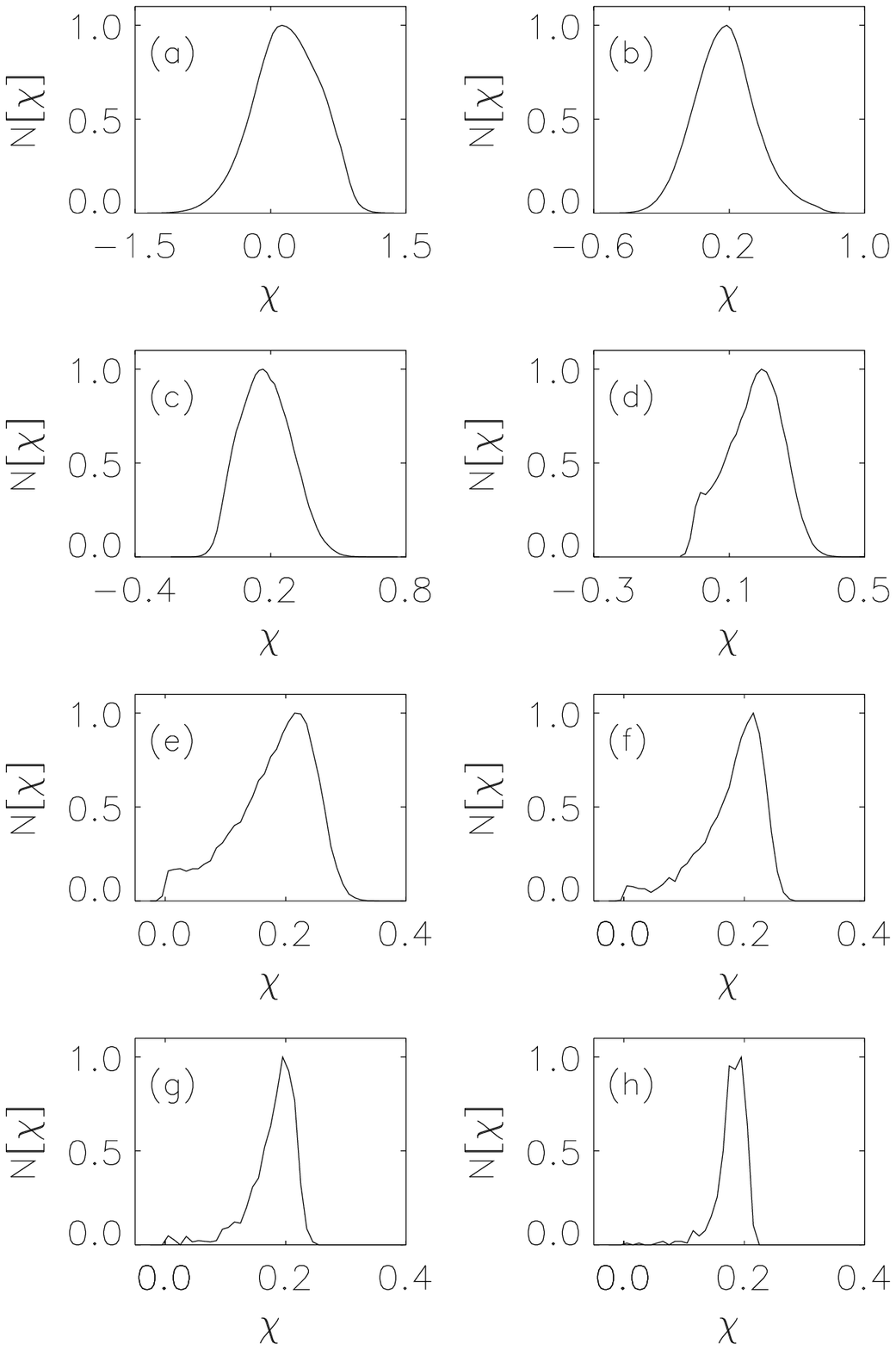}
           }
        \begin{minipage}{15cm}
        \end{minipage}
        \vskip -0.0in\hskip -0.0in
\caption{ $N[{\chi}(t)]$, the distribution of short time Lyapunov exponents
for a collection of chaotic orbits with energy $E=6$ evolved in the potential
(2) with $a=b=1$, computed as a function of the sampling time $t$.
(a) $t=1$. (b) $t=4$. (c) $t=16$. (d) $t=64$. (e) $t=256$. (f) $t=1024$.
(g) $t=4096$ (h) $t=16384$.}
\vspace{0.0cm}
\end{figure}
\vfill\eject

\pagestyle{empty}
\begin{figure}[t]
\centering
\centerline{
        \epsfxsize=8cm
        \epsffile{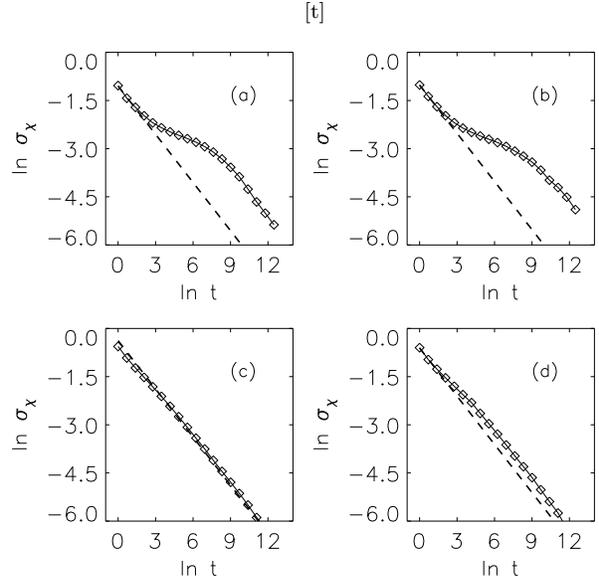}
           }
        \begin{minipage}{15cm}
        \end{minipage}
        \vskip -0.0in\hskip -0.0in
\caption{ (a) The dispersion ${\sigma}_{\chi}(t)$, plotted as a function
of the sampling time $t$, for a collection of chaotic orbits with energy
$E=4$ evolved in the potential (2) with $a=b=1$. The dashed line corresponds
to a $t^{-1/2}$ dependence. (b) The same for chaotic
orbits with $E=6$ and $a=b=1$. (c) The same for chaotic orbits with $E=6$
and $a=3$ and $b=-1$. (d) The same for chaotic orbits with $E=10$
and $a=3$ and $b=-1$.}
\vspace{-0.0cm}
\end{figure}

\pagestyle{empty}
\begin{figure}[t]
\centering
\centerline{
        \epsfxsize=8cm
        \epsffile{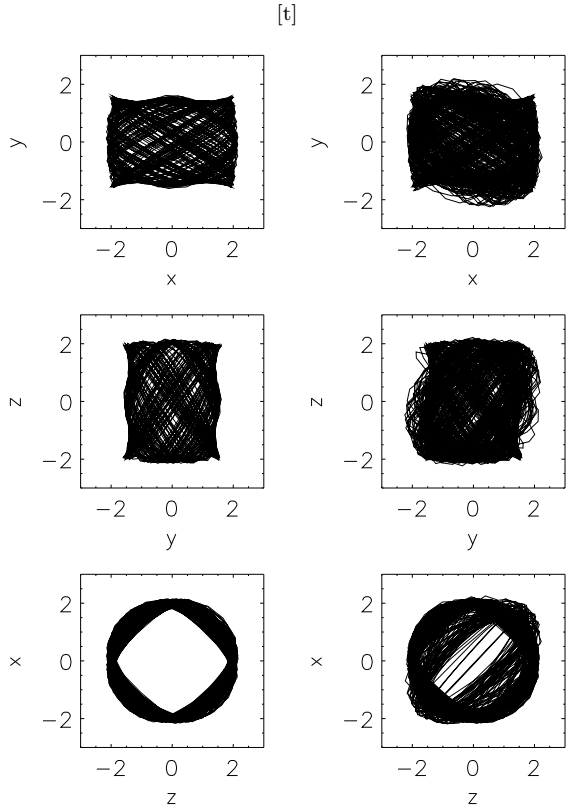}
           }
        \begin{minipage}{15cm}
        \end{minipage}
        \vskip -0.0in\hskip -0.0in
\caption{ Visual representations of the orbit used to compute Fig. 4.
The left hand panels exhibit data for times $15700<t<16212$. The right hand
panels exhibit data for $15700<t<16724$, an interval twice as long.}
\end{figure}

\pagestyle{empty}
\begin{figure}[t]
\centering
\centerline{
        \epsfxsize=8cm
        \epsffile{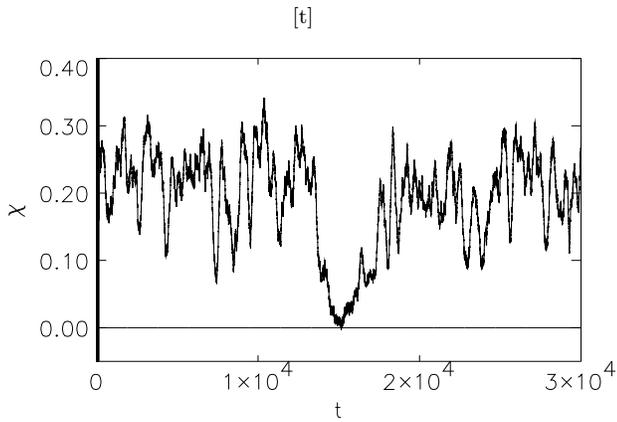}
           }
        \begin{minipage}{15cm}
        \end{minipage}
        \vskip -2.0in\hskip -0.0in
\caption{ A coarse-grained estimate of the largest short Lyapunov exponent
${\chi}$, generated for a chaotic orbit in the potential (3) with $E=4$ and 
$a=b=1$. The coarse-graining was effected by computing ${\chi}(t)$ for 
successive intervals ${\delta}t=1$, and performing a box-car average over 
$256$ adjacent intervals.}
\vspace{-5.0cm}
\end{figure}

\pagestyle{empty}
\begin{figure}[t]
\centering
\centerline{
        \epsfxsize=8cm
        \epsffile{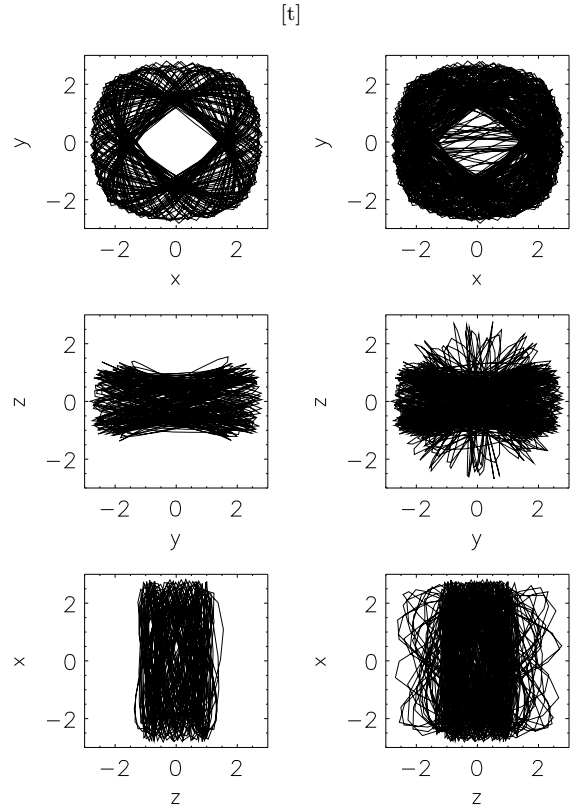}
           }
        \begin{minipage}{15cm}
        \end{minipage}
        \vskip -0.0in\hskip -0.0in
\caption{ The analogue of Fig. 5 for an initial condition with $E=10$, in
this case integrated in the potential (3) with $a=1.6$ and $b=0.4$. In this
case, the integration times for the left and right hand panels are 
${\Delta}t=400$ and ${\Delta}t=800$.}
\end{figure}

\pagestyle{empty}
\begin{figure}[t]
\centering
\centerline{
        \epsfxsize=8cm
        \epsffile{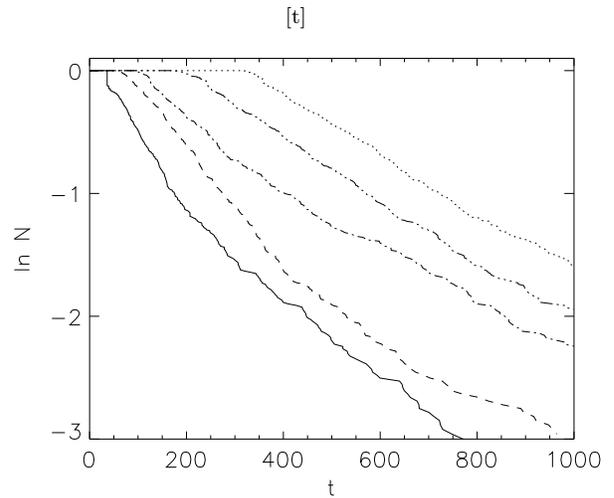}
           }
        \begin{minipage}{15cm}
        \end{minipage}
        \vskip -0.0in\hskip -0.0in
\caption{ $N(t)$, the fraction of the orbits from a 1000 orbit ensemble not 
yet having escaped at time $t$, computed for an initial condition with $E=10.0$
evolved in the dihedral potential with $a=0.4$ and $b=1.6$ in the presence 
of periodic driving with $5{\;}{\le}{\;}{\omega}{\;}{\le}{\;}6$ and variable
amplitude $A=10^{-2}$ (solid line), $A=10^{-3}$ (dashed), $A=10^{-4}$ 
(dot-dashed), $A=10^{-5}$ (triple-dot-dashed), and $A=10^{-7}$ (dotted).}
\vspace{-5.0cm}
\end{figure}

\pagestyle{empty}
\begin{figure}[t]
\centering
\centerline{
        \epsfxsize=8cm
        \epsffile{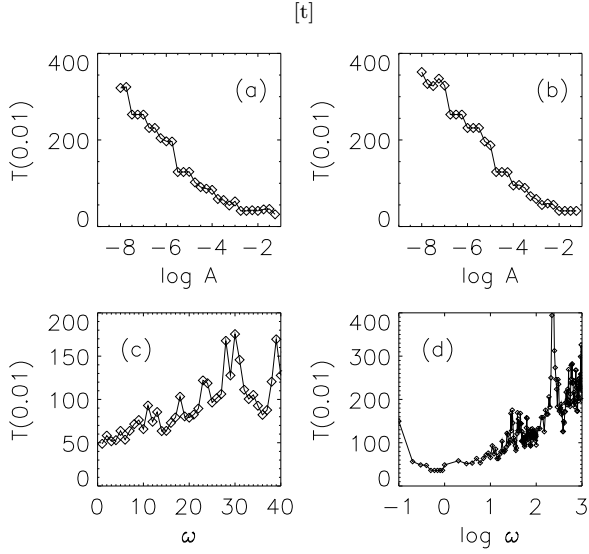}
           }
        \begin{minipage}{15cm}
        \end{minipage}
        \vskip -0.8in\hskip -0.0in
\caption{(a) $T(0.01)$, the first escape time for 1\% of an ensemble of 
2000 integrations of the initial condition in Fig. 6, driven with frequencies  
$2{\;}{\le}{\;}{\omega}{\;}{\le}{\;}3$ and variable amplitude. (b) The same
for $5{\;}{\le}{\;}{\omega}{\;}{\le}{\;}6$. (c) $T(0.01)$ for the same 
initial condition, now plotted as a function of ${\omega}$ for fixed amplitude
$A=10^{-3}$. (d) $T(0.01)$ plotted as a function of ${\rm log}\,{\omega}$.
}
\vspace{0.0cm}
\end{figure}

\pagestyle{empty}
\begin{figure}[t]
\centering
\centerline{
        \epsfxsize=8cm
        \epsffile{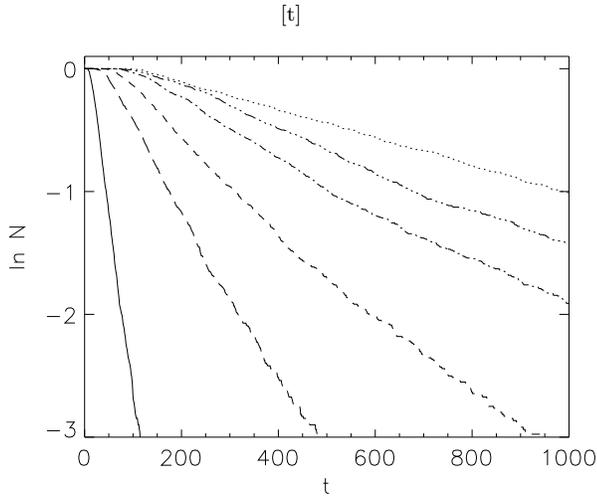}
           }
        \begin{minipage}{15cm}
        \end{minipage}
        \vskip -0.0in\hskip -0.0in
\caption{ $N(t)$, the fraction of the orbits from a 1000 orbit ensemble not 
yet having escaped at time $t$, computed for an initial condition with $E=10.0$
evolved in the dihedral potential with $a=0.4$ and $b=1.6$ in the presence 
of additive white noise with ${\Theta}=10.0$ and variable ${\eta}=10^{-3}$ 
(solid line), ${\eta}=10^{-4}$ (broad dashes), ${\eta}=10^{-4.5}$ (short 
dashes), ${\eta}=10^{-5}$  (dot-dashed), ${\eta}=10^{-5.5}$ 
(triple-dot-dashed), and ${\eta}=10^{-8}$ (dotted).}
\vspace{0.0cm}
\end{figure}
\vfill\eject
\pagestyle{empty}
\begin{figure}[t]
\centering
\centerline{
        \epsfxsize=8cm
        \epsffile{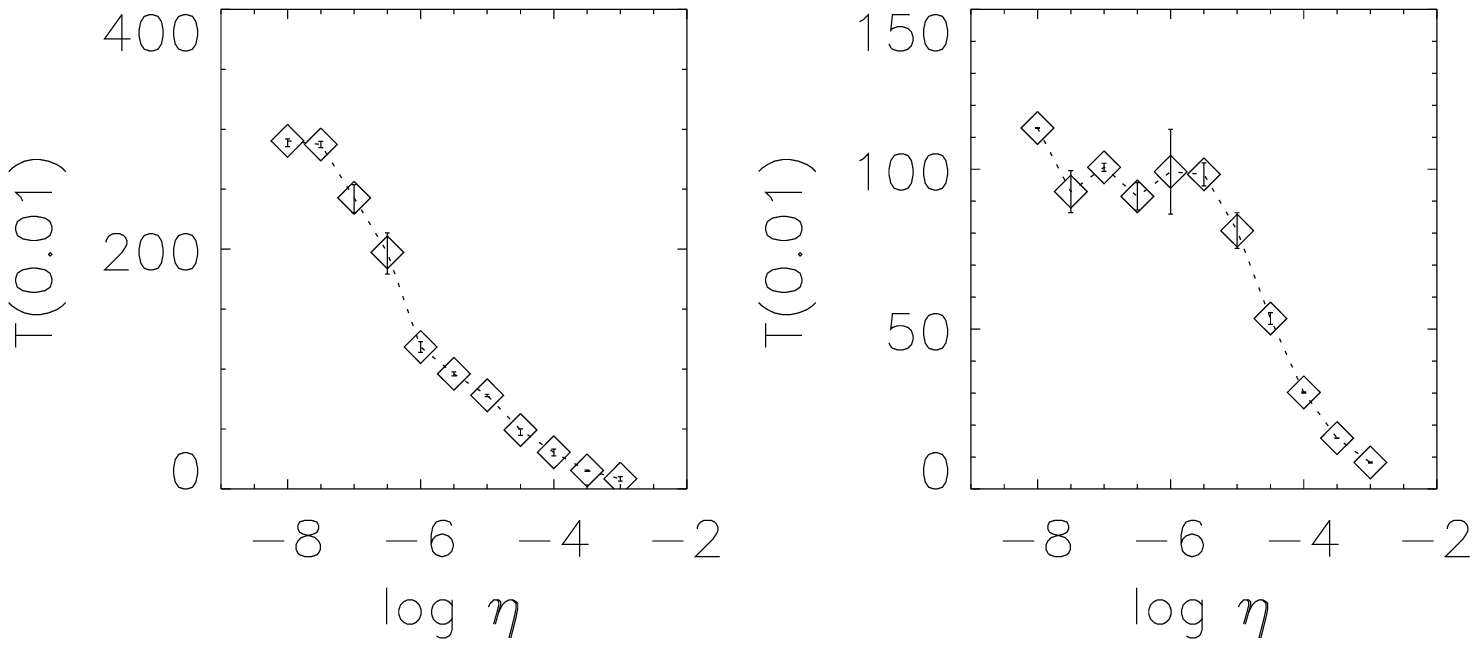}
           }
        \begin{minipage}{15cm}
        \end{minipage}
        \vskip 0.4in\hskip -0.0in
\caption{(a) $T(0.01)$, the first escape time for 1\% of an ensemble of 
2000 integrations of the initial condition in Fig. 6, perturbed by additive
white noise with ${\Theta}=10.0$ and variable ${\eta}$. (b) The same for a 
different initial condition.}
\vspace{0.0cm}
\end{figure}

\pagestyle{empty}
\begin{figure}[t]
\centering
\centerline{
        \epsfxsize=8cm
        \epsffile{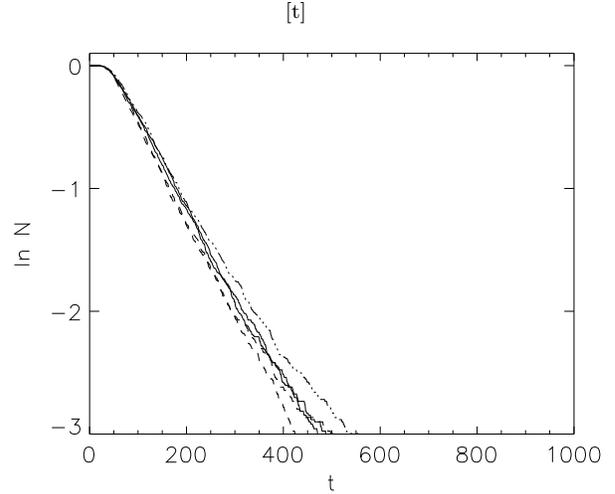}
           }
        \begin{minipage}{15cm}
        \end{minipage}
        \vskip -0.0in\hskip -0.0in
\caption{ $N(t)$, the fraction of the orbits from a 1000 orbit ensemble not 
yet having escaped at time $t$, computed for an initial condition with $E=10.0$
evolved in the dihedral potential with $a=0.4$ and $b=1.6$ in the presence 
of different white noises with ${\Theta}=10.0$ and ${\eta}_{0}=10^{-4}$. The
solid curve represents additive white noise and friction with constant 
${\eta}$. The dot-dashed, triple-dot-dashed, and dotted curves represent, 
respectively, friction and noise with ${\eta}{\;}{\propto}{\;}v$, $v^{2}$ and 
$v^{3}$. The dashed curve represents additive white noise but vanishing 
friction.
}
\vspace{0.0cm}
\end{figure}

\pagestyle{empty}
\begin{figure}[t]
\centering
\centerline{
        \epsfxsize=8cm
        \epsffile{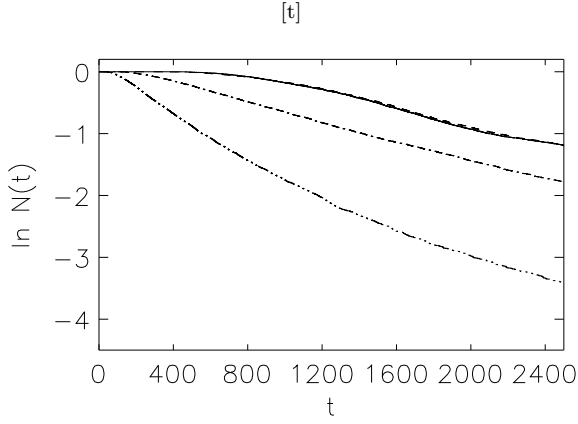}
           }
        \begin{minipage}{15cm}
        \end{minipage}
        \vskip -1.8in\hskip -0.0in
\caption{$N(t)$, the fraction of the orbits from a $4800$ orbit ensemble not 
yet having escaped at time $t$, computed for the initial condition exhibited 
in Fig. 9, allowing for friction and colored noise given by eq. (5.2) with
${\Theta}=10$, ${\eta}=10^{-5}$, and either white noise (triple-dot-dashed) or 
variable ${\alpha}=2.0$ (dot-dashed), ${\alpha}=0.2$ (broad dashes), and 
${\alpha}=0.02$ (solid).}
\vspace{-0.2cm}
\end{figure}

\pagestyle{empty}
\begin{figure}[t]
\centering
\centerline{
        \epsfxsize=8cm
        \epsffile{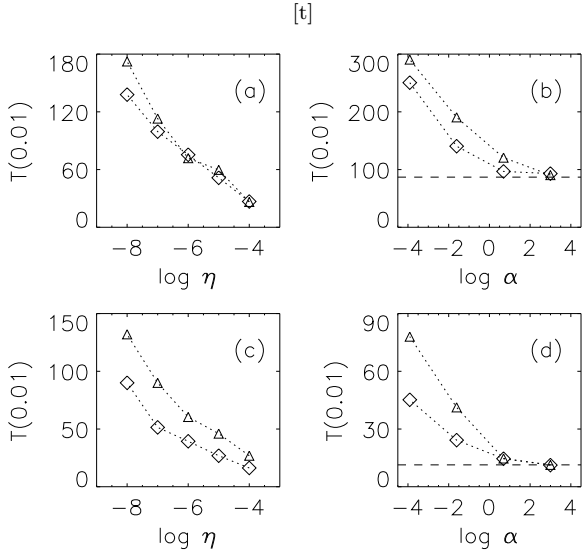}
           }
        \begin{minipage}{15cm}
        \end{minipage}
        \vskip -1.5in\hskip -0.0in
\caption{(a) $T(0.01)$, the first escape time for 1\% of an ensemble of $4800$
integrations, computed for the initial condition used to generate Fig. 9,
plotted as a function of ${\rm log}\,{\eta}$ for fixed ${\alpha}=2.0$ for the 
stochastic processes defined by (5.1) (diamonds) and (5.2) (triangles), 
allowing for both friction and noise. (b) $T(0.01)$ for the same initial
condition, plotted as a function of ${\rm log}\,{\alpha}$ for fixed 
${\eta}=10^{-5}$, for the stochastic processes (5.1) (diamonds) and (5.2) 
(triangles), again allowing for both friction and noise. The dashed line
represents the asymptotic value for white noise $({\alpha}\to\infty)$.
(c) The same as (a), albeit for a different initial condition and with
${\alpha}=0.2$. (d) The same as (b), albeit for the initial condition in (c)
and with ${\eta}=10^{-7}$.
as (a) and (b) for another initial condition. }
\vspace{0.0cm}
\end{figure}

\pagestyle{empty}
\begin{figure}[t]
\centering
\centerline{
        \epsfxsize=8cm
        \epsffile{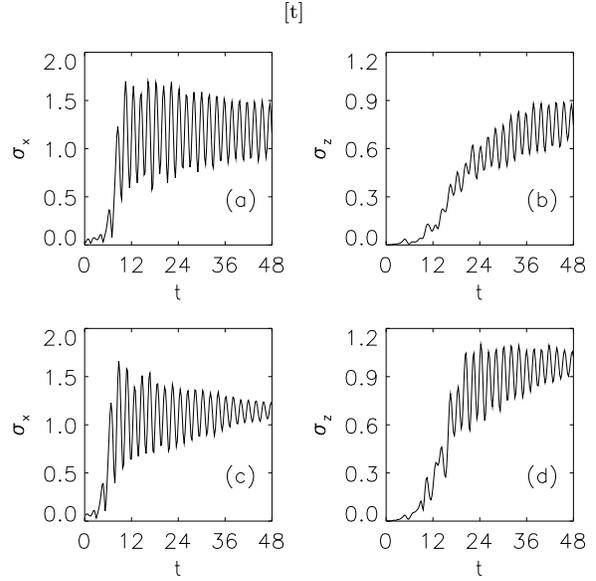}
           }
        \begin{minipage}{15cm}
        \end{minipage}
        \vskip -0.0in\hskip -0.0in
\caption{(a) The dispersion ${\sigma}_{x}$ generated for a localised ensemble
of $6400$ initial conditions with $E=10$ evolved without 
perturbations in the potential (2) with $a=1=1$. (b) The dispersion 
${\sigma}_{z}$ for the same ensemble. (c) ${\sigma}_{x}$ for the same ensemble,
now allowing for periodic driving with ${\Omega}=1$ and $A=10^{-1.5}$. (d) 
${\sigma}_{x}$ for the same ensemble, allowing for periodic driving with 
${\Omega}=1$ and $A=10^{-1.5}$. }
\vspace{0.0cm}
\end{figure}

\pagestyle{empty}
\begin{figure}[t]
\centering
\centerline{
        \epsfxsize=8cm
        \epsffile{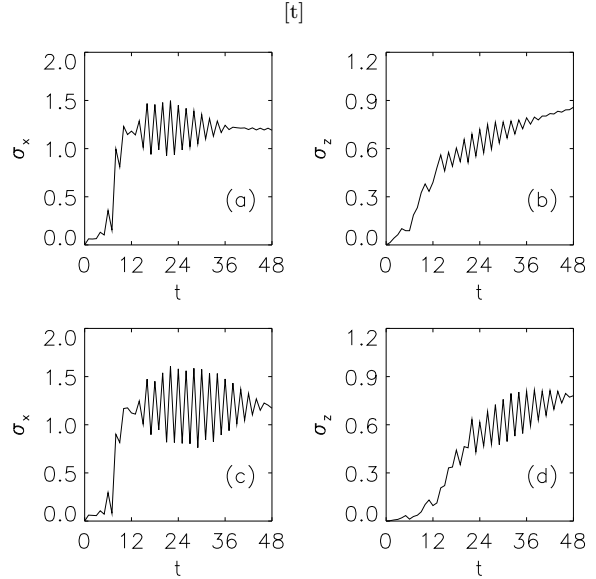}
           }
        \begin{minipage}{15cm}
        \end{minipage}
        \vskip -0.0in\hskip -0.0in
\caption{(a) The dispersion ${\sigma}_{x}$ generated for the same ensemble
evolved in the presence of friction and additive white noise with ${\Theta}=10$
and ${\eta}=10^{-4}$. (b) The dispersion ${\sigma}_{z}$ for the same 
computation. (c) The same as (a) but with ${\eta}=10^{-6}$. (d) The same as (b)
but with ${\eta}=10^{-6}$.}
\vspace{0.0cm}
\end{figure}

\pagestyle{empty}
\begin{figure}[t]
\centering
\centerline{
        \epsfxsize=8cm
        \epsffile{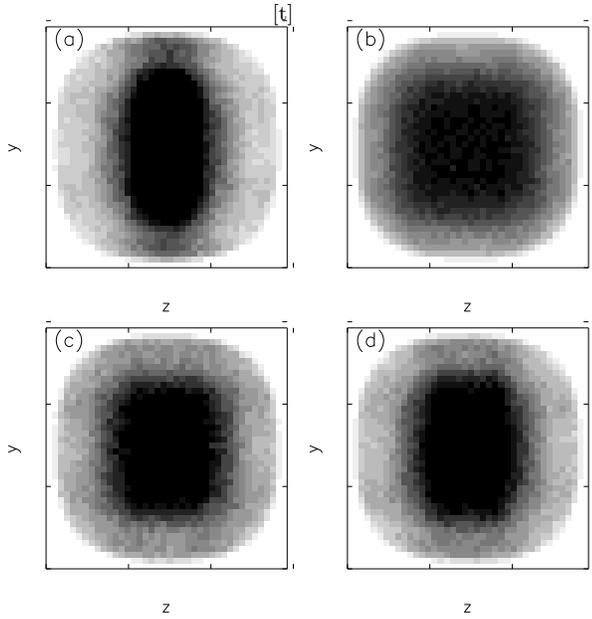}
           }
        \begin{minipage}{15cm}
        \end{minipage}
        \vskip -0.0in\hskip -0.0in
\caption{(a) Gray scale plot of the distribution $f(x,y)$ generated for a 
localised ensemble of $6400$ initial conditions with $E=10$ evolved without 
perturbations in the potential (2) with $a=b=1$, with data recorded at 
intervals ${\delta}t=1$ for  $64<t<128$. (b) The invariant distribution 
associated with a much longer time integration. (c) The same as (a) but 
allowing for periodic driving with ${\Omega}=1$ and $A=10^{-1.5}$. (d) The
same as (c) but with $A=10^{-2.5}$.}
\vspace{0.0cm}
\end{figure}

\pagestyle{empty}
\begin{figure}[t]
\centering
\centerline{
        \epsfxsize=8cm
        \epsffile{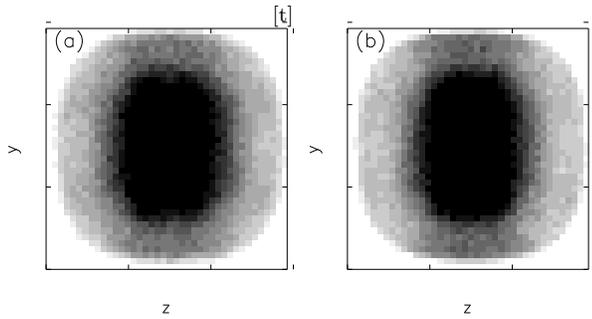}
           }
        \begin{minipage}{15cm}
        \end{minipage}
        \vskip -1.5in\hskip -0.0in
\caption{(a) Gray scale plot of the distribution $f(x,y)$ generated for the
same initial ensembles as in the preceding figure, now allowing for additive
white noise with ${\Theta}=10$ and ${\eta}=10^{-4}$. (b) The same as (a) but
with ${\eta}=10^{-6}$.}
\vspace{0.0cm}
\end{figure}

\pagestyle{empty}
\begin{figure}[t]
\centering
\centerline{
        \epsfxsize=8cm
        \epsffile{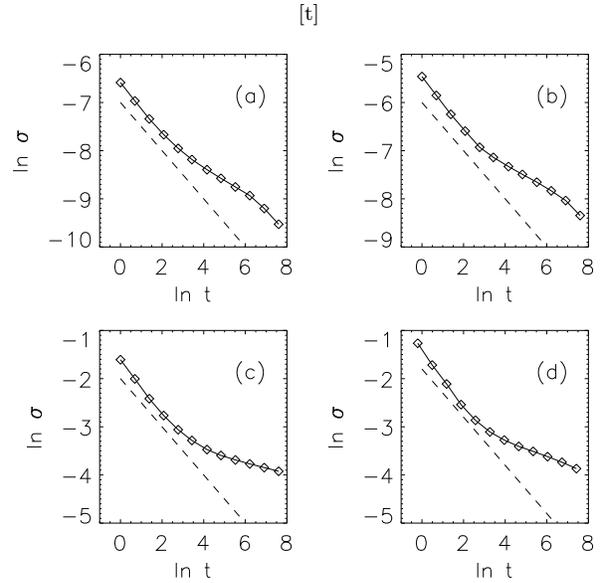}
           }
        \begin{minipage}{15cm}
        \end{minipage}
        \vskip -0.0in\hskip -0.0in
\caption{The analogue of Fig. 3 for chaotic orbits in the maximally traxial
${\gamma}=1$ Dehnen potential. The upper panels are for the two highest
energies used by Merritt \& Fridman (1996). The lower panels are for the
two lowest energies.}
\vspace{0.0cm}
\end{figure}

\appendix

\label{lastpage}
\end{document}